\documentclass[rmp,nobibnotes,floats]{revtex4}
\usepackage{amsmath}
\usepackage{graphicx}
\bibpunct{(}{)}{;}{a}{,}{,}

\begin{document}

\title{Steady-state mode III cracks in a viscoelastic 
lattice model}

\author{Leonid Pechenik and  Herbert Levine}
\affiliation{Department of Physics\ \ \ \ \ \ \ \ \ \ \ \ \ \ \ \ \ \ \ \ \ \ \ \ \ \ \ \ \ \ \ \ \ \ \ \ \ \ \ \ \ \ \ \\
University of California, San Diego\ \ \ \ \ \ \ \ \ \ \ \ \ \ \ \ \ \ \ \ \ \ \ \ \ \ \ \ \ \ \ \ \ \ \ \ \ \ \ \ \ \ \ \\
La Jolla, CA  92093-0319 USA}
\author{David A. Kessler}
\thanks{Corresponding Author;  FAX: +972-3-535-3298}
\email{kessler@dave.ph.biu.ac.il}
\affiliation{Deptartment of Physics\ \ \ \ \ \ \ \ \ \ \ \ \ \ \ \ \ \ \ \ \ \ \ \ \ \ \ \ \ \ \ \ \ \ \ \ \ \ \ \ \ \ \ \\
Bar-Ilan University\ \ \ \ \ \ \ \ \ \ \ \ \ \ \ \ \ \ \ \ \ \ \ \ \ \ \ \ \ \ \ \ \ \ \ \ \ \ \ \ \ \ \ \ \ \ \ \ \ \\\
Ramat-Gan, Israel}

\begin{abstract}
We extend the Slepyan  solution of the problem of a steady-state crack in 
an infinite ideally brittle lattice model to include dissipation
in the form of Kelvin viscosity.  As a demonstration of this technique,
based on the Wiener-Hopf method, we apply the method to mode III cracks 
in a square lattice.  We use this solution to find the critical
velocity at which the steady-state solution becomes inconsistent due
to additional bond-breaking; this point signaling the onset of complex
dynamical behavior.\newline\newline
\end{abstract}
\maketitle

\section{Introduction}
Recent experiments on dynamical fracture \citep{exp1,exp2} have shown
that cracks in brittle materials become unstable above a critical
velocity. This instability is associated with the formation
of a roughened fracture surface, with an increased dissipation of elastic
energy, and with complicated velocity dynamics - for a review,
see \citep{review}. Although there are hints of such an instability in
the traditional continuum formulations of ideally brittle cracks 
\citep{yoffe}, a
systematic treatment does not appear possible. Indeed, recent attempts
\citep{langer-new} 
to utilize a cohesive stress modification \citep{barenblatt} of the 
continuum elastic
equations in order to address this problem have been rather unsuccessful.

In the absence of a compelling continuum approach, one must deal in
some manner with discrete dynamics on the ``atomic scale". In this
regard, Slepyan \citep{slepyan,slepyan3,slepyan2} pioneered the 
idea of studying lattice models in
which atoms interact (via piece-wise linear springs) with their neighbors in 
a predetermined lattice geometry. This conceptual framework was
further developed by Marder and collaborators \citep{marder_gross,marder-liu}.
Here, one can directly obtain the relationship between the crack tip
velocity and the imposed driving and furthermore one can search for
conditions which do not allow stable steady crack motion. Clearly,
lattice models are not fully realistic, especially at displacements that
are not small compared to the underlying lattice spacing; to be more
realistic, one must resort to molecular dynamics simulations including
all possible atomic interactions \citep{md0,md1,md1a,md2,md2a,md3}. However, the analytic
tractability of these models as well as 
indications \citep{marder_gross,sander1,sander2} that they do contain
the essential mechanism responsible for the observed dynamical instability
make them well worthy of serious attention.

Slepyan recognized that in
an ideally brittle material (where each spring is linear until a displacement at
which it completely breaks) one can use Fourier methods to find
the lattice analog of a traveling wave solution. In this solution, each
point on the lattice (at the same transverse coordinate) undergoes the
same time history of motion as any other, albeit with some time delay.
Once one obtains the solution, one must check that all springs assumed
to be linear have displacements that are in fact below the breaking threshold.
The violation of this assumption by the steadily propagating solution
signals the onset of more complex dynamical behavior, at least in
qualitative accord to what is seen experimentally \citep{review,marder_gross}.
Using the Wiener-Hopf technique, Slepyan solved for the velocity-driving
curve in the limit  where the width of the lattice transverse to the 
crack direction  goes to infinity.

One important consideration in all models of brittle fracture concerns
dissipation mechanisms. From our perspective, it makes sense to put in
dissipation at the lattice scale in such a way so as not dominate
the large-scale continuum elastic field. If in addition one demands
an equation that is local in time, one is led \citep{langer,sander1}  
to the introduction
of a Kelvin viscosity which dissipates energy proportional to
the rate of change of spring lengths. In the naive continuum limit,
this gives rise to a third derivative (two space, one time) term. That
means that if the viscosity is chosen to be O(1) on the lattice scale,
it will scale to zero as far as the macroscopic dynamics is concerned.
We saw this explicitly in a previous set of papers \citep{kl1,dk} which
solved this problem for finite width lattices and considered the
nature of the solution as the width became large. Nevertheless, the viscosity  can
have a considerable effect on aspects of the solution which depend on
the microscopic details, namely the crack speed (for a fixed stress
intensity factor) and the self-consistency of the traveling wave ansatz.

The rest of this paper is organized as follows. In the next section,
we formally define the model, employing Slepyan's idea \citep{slepyan,slepyan2} 
of replacing the driving at some external boundary with some local
forcing on the crack surface. This allows for the solution in Section III of 
the strictly infinite lattice model; the connection to a finite problem
with some displacement driving is made by appealing to the universality
of the microscopic crack solution given a fixed stress intensity factor.
In the following section, we discuss the numerical evaluation
of the  velocity curve from its formal expression. Following that, we
turn to the aforementioned self-consistency condition and study the
effect of dissipation on the critical velocity. We conclude with some
observations in the final section.

\section{Square lattice model}

We wish to study mode III cracks and the effects thereupon of Kelvin
viscosity. Our model thus consists of a square lattice of mass points
undergoing (scalar) displacements out of the plane. The lattice extends
infinitely long in both the $x$ (along-crack) and $y$ (transverse) directions.
The lattice points are connected to their nearest neighbors by ideally brittle
springs which
behave linearly with spring constant 1 until some threshold elongation
$2 \epsilon$ at which point they irreversibly crack. All the
(uncracked) springs have a viscous damping $\eta$.  
Let us assume
that the crack corresponds to a sequential breaking of
the vertical bonds between the mass points at rows $y=1$ and
$y=0$. The equation of motion for the masses in rows $y>1$ reads
\begin{equation}
\label{ynot1}
{\ddot u} (x,y) = \left(1 + \eta \frac{d}{dt}\right) \Big(u(x+1,y) +
u(x-1,y) +
u(x,y+1) + u(x,y-1) - 4u(x,y)  \Big) \ .
\end{equation}
We allow the vertical bonds along the crack surface
to have a different spring constant $k$ and damping parameter $\tilde{\eta}$,
to model the effect of having weak  links  along the crack surface. The
equation for $y=1$ then reads, using the assumed symmetry $u(x,y)=-u(x,1-y)$,
\begin{eqnarray}
{\ddot u}(x,y) &=& \left(1 + \eta \frac{\partial}{\partial t}\right)\Big(u(x+1,1)+u(x-1,1)+
u(x,2) - 3 u(x,1)\Big)\nonumber \\ 
 &\ &\quad + \theta(\epsilon-u(x,1))\left(k+\tilde{\eta}\frac{\partial}{\partial t}\right)\Big( -2u(x,1)\Big)
 \ .
\end{eqnarray}
It is easily verified that the equations for $y \le 0$ are consistent with
the above symmetry.
We can, without loss of generality choose $\epsilon=1/2$
(so that the threshold elongation is unity). 
Note that in these units, the elastic wave speed is unity, so all
velocities are dimensionless, expressed as fractions of the wave speed.

We are interested in steady-state cracks, described by the Slepyan
traveling wave ansatz,
\begin{equation}
u(x,y,t) = u(x-vt,y)
\end{equation}
which implies that every mass point in a given row undergoes the
same time history, translated in time.  We choose the origin of
time such that $u(0,1)=\epsilon$, so that it represents the moment
of cracking of the spring in column 0.
If we define $\tau=x-vt$,
the equations of motion for $y>1$ become
\begin{eqnarray}\label{eomy}
\left(1 - \eta v \frac{\partial}{\partial \tau}\right)  
\Big(u(\tau+1,y)&+&u(\tau-1,y) +  u(\tau,y-1) 
  \nonumber \\ &+& u(\tau,y-1)-4 u(\tau,y)\Big) 
-v^2\frac{\partial^2 u(\tau,y)}{\partial \tau^2}\ = 0 \ .
\end{eqnarray}
For $y=1$, we separate out the terms proportional to $\theta(-\tau)$, giving
\begin{eqnarray}\label{eom1}
\left(1 - \eta v \frac{\partial}{\partial\tau}\right)\Big(u(\tau+1,1)+u(\tau-1,1)&+&
u(\tau,2) - 3 u(\tau,1)\Big)\nonumber \\ 
 + \left(k-\tilde{\eta}v\frac{\partial}{\partial\tau}\right)\Big( -2u(\tau,1)
\Big )
&-&v^2\frac{\partial^2u(\tau,1)}{\partial\tau^2}
= -\sigma(\tau) \ .
\end{eqnarray}
The driving term for this equation has the form
\begin{equation}\label{forcek}
\sigma(\tau)=\theta(-\tau)\left[F_0+2\left(k-\tilde{\eta} v
\frac{\partial}{\partial \tau}\right)u(\tau,1)\right] \ .
\end{equation}
In this
expression, we have inserted an external force $F_0$ which acts
on the crack boundaries. This force is an artifice which serves
as an inhomogeneous source term, in place of introducing the driving through
the boundary condition on the top and bottom surfaces.
We will see later  how to choose this function.

We Fourier transform with respect to $\tau$ (see the appendix for our conventions regarding Fourier transforms
and Fourier integrals).
If we assume 
that $u(q,y) =  \chi_q [\xi(q)]^{y-1} $ for $y \ge 1$,
we obtain from Eqn. (\ref{eomy}) the dispersion relationship
\begin{equation}
(1 + i q \eta v)(e^{- i q}+ e^{ i q} + \xi + \frac{1}{\xi}-4)+q^2 v^2= 0 
\end{equation}
This can be rewritten as 
\begin{equation}
(\xi^{1/2}-\frac{1}{\xi^{1/2}})^2=4 \sin^2 \frac{q}{2}-\frac{q^2 v^2} {1 + i q \eta v}\equiv h^2
\end{equation}
which therefore implies that $\xi^{1/2}=\pm h/2 \pm \sqrt{(h/2)^2+1}$.
If we define $$ r^2\equiv h^2+4$$
then we can write
\begin{equation}\label{xi}
\xi=\frac{(r-h)^2}{4}=1-\frac{h(h-r)}{2}=\frac{r-h}{r+h}
\end{equation}
Note that we have must choose the sign that corresponds to a decaying
solution in the $y$ direction, i.e. $|\xi|<1$.

If we substitute this ansatz into Eqn. (\ref{eom1}), we get
\begin{equation}\label{chi}
\left(-2 ( k + i q \tilde{\eta} v) + 
(1 + i q \eta v) ( 1-\frac{1}{\xi})\right)\chi=-\sigma^F
\end{equation}
where $\sigma^F$ is the Fourier transform of $\sigma$.
Noting that $-1-\frac{1}{\xi}=-\frac{r(r+h)}{2}$, we obtain
\begin{equation}
\chi_q=\frac{2 \sigma^F}{N(q)}
\end{equation}
with
\begin{equation}
N(q)=h(r+h)(1+ i \eta v q)+ 4 (k + i \tilde{\eta} v q ) \ .
\end{equation}

This equation is implicit, since the driving term $\sigma^F$ on the
right hand side still depends on the displacement field $u(\tau,1)$. In
the next section, we solve this equation via the Wiener-Hopf technique.

\section{Wiener-Hopf solution}

Let us define
$\vee(\tau)=2u(\tau,1)$, the bond elongation between rows $y=0$ and $y=1$. 
In the appendix, we define $\vee^+ (q) $ and $\vee^- (q)$ as the decomposition
of $\vee(\tau)$ into terms which are analytic in the upper and lower
half planes respectively. This allows us to write Eq. (\ref{forcek}) as
\begin{equation}
\sigma^F(q)=F_0^-(q)+ (k+i \tilde{\eta} v q) \vee^-(q) -  \tilde{\eta} v
\vee(0)
\end{equation}
where we have employed a similar breakup for the driving term $F_0$,
and where we have utilized Eq. (\ref{derivative}). 

Substituting the formula for $ \sigma^F$ into Eqn. (\ref{chi})
for $\chi=(\vee^+ + \vee^-)/2$
leads to a closed form equation for $\vee$; namely
\begin{equation}\label{rawk}
\vee^++\vee^-(1-\frac{4(k+i \tilde{\eta} v q)}{N(q)})
=\frac{4 F_0^-}{N(q)}-
\frac{4 \tilde{\eta} v \vee (0)}{N(q)} \ .
\end{equation}
Defining $S$ by
\begin{equation}
\label{seqn}
S\equiv 1-\frac{4(k+i \tilde{\eta} v q)}{N(q)}
=\frac{h(1+i\eta v q)}{h(1+ i \eta v q)+ (r-h) (k + i \tilde{\eta} v q)} \ ,
\end{equation}
we can rewrite Eq. (\ref{rawk}) as
\begin{equation}\label{s1k}
\vee^++S\vee^-+\frac{ \tilde{\eta} v }{k + i q \tilde{\eta} v} 
(1-S) \vee (0)
=\frac{(1-S) F_0^-}{k + i q \tilde{\eta} v} \ .
\end{equation}

We will solve this equation using the Wiener-Hopf technique. The
key is to factor $S$ into a product of two pieces, $S(q)=S^+(q)S^-(q)$,
each of which is regular
in the upper ($S^+$) or lower ($S^-$) half plane. Doing so, we
get
\begin{equation}\label{mainpsp}
\frac{\vee^+}{S^+}+S^-\vee^-+\frac{ \eta v }{k + i q \tilde{\eta} v} 
\left (\frac{1}{S^+}-S^- \right ) \vee (0)=\frac{1-S}{S^+} 
\frac{F_0^-}{k + i q \tilde{\eta} v} \ .
\end{equation}
We then have to deal with the extra singularity at $q=\frac{ik}{\eta v}$. We
do this by a simple subtraction
\begin{equation}  
\frac{\vee^+}{S^+}+S^-\vee^-+\frac{ \tilde{\eta} v \vee (0)}{k + i q \tilde{\eta} v} 
\left(\frac{1}{S^+}-
\left. \frac{1}{S^+} \right| _{q=\frac{ik}{\tilde{\eta} v}}+\left. \frac{1}{S^+}
\right|_{q=\frac{ik}{\tilde{\eta} v}}-S^- \right) \ =\  \frac{1-S}{S^+} 
\frac{F_0^-}{k + i q \tilde{\eta} v} \ .
\end{equation} 
We have now to choose $F_0$, our external forcing.  In taking the
width to infinity, in essence we are solving for the ``inner'' solution
in the sense of boundary-layer theory; i.e., only on the scale of the
lattice spacing.  The external forcing in the finite width
problem only acts on the large scale, and does not vary on the lattice
scale.  Thus, $F_0$ must have support only at $q=0$.
Following Slepyan, then, we choose $F_0$ such that
\begin{equation}\label{defBk}
\frac{1-S}
{S^+ (k + i q \tilde{\eta} v) }F_0^-=2\pi B \delta (q)
=B \left( \frac{1}{iq+0}+\frac{1}{-iq+0}\right) \ ,
\end{equation}
where $B$ is a constant.
This then gives us the the two separate equations
\begin{subequations}\label{vpm}
\begin{eqnarray}\label{vp}
\vee^+&=&-\frac{ {\tilde\eta} v \vee (0)}{k + i q \tilde{\eta} v}
\left(1-
\frac{S^+}{S^+| _{q=\frac{ik}{\tilde{\eta} v}}}\right) + \frac{BS^+}{-iq+0}\ , \\
\label{vm}
\vee^-&=&-\frac{ \tilde{\eta} v \vee (0)}{k + i q \tilde{\eta} v}\left( \frac{1}{S^-S^+|_{q=\frac{ik}{\eta v}}}-1 \right) + \frac{B}{(iq+0)S^-} \ .
\end{eqnarray}
\end{subequations}
As shown in Eqs. (\ref{limp}) and (\ref{limm}) we can solve for $\vee (0)$
from Eqs. (\ref{vp}) or (\ref{vm}) by multiplying the first by $ - i q \rightarrow + 
\infty $ or the second by $ i q \rightarrow + \infty $. This yields
\begin{equation}\label{v0bk}
\vee (0)= B \left. S^+\right|_{ q=\frac{i k}{\tilde{\eta}
v} } \ .
\end{equation}

Substituting this back into Eqns. (\ref{vpm}), and defining $\eta_k\equiv
{\tilde \eta}/k$, we find
\begin{subequations}
\setlength{\jot}{20pt}
\begin{eqnarray}
\vee^+&=&\frac{B S^+}{(-iq+0)(1 + i q \eta_k v)}  
-\frac{ \eta_k v B}{1
+ i q \eta_k v}\left.S^+\right|_{q=\frac{i}{\eta_k v}} \ ,\\
\vee^-&=&  \frac{B}{(iq+0)(1 + i q \eta_k v)S^-}
+\frac{ \eta_k v B}{1 + i q \eta_k v}
\left.S^+\right| _{q=\frac{i}{\eta_k v}}S^-
\end{eqnarray}
\end{subequations}

These equations directly determine the displacements given the strength
of the external driving, $B$. To get a physical handle on the meaning of this
constant, we note that Eqns. (\ref{vpm}) can be directly solved
for the leading behavior of $\vee^{\pm}$ as $q$ goes to $0$,
using the fact that in this limit,
\begin{equation}\label{kappa}
S^{\pm} \sim \kappa^{\pm 1}\sqrt{A}(q \pm i 0)^{1/2}
\end{equation}
where $A=\frac{\sqrt{1-v^2}}{2k}$, and $\kappa$ is a constant,
reflecting the arbitrariness in how we perform the decomposition of $S$.
We find
\begin{subequations}
\begin{eqnarray}
\vee_+ \sim \kappa\sqrt{A}\frac{i^{1/2} B}{(-iq+0)^{1/2}}\\
\vee_- \sim \frac{\kappa}{\sqrt{A}}\frac{i^{1/2} B}{(iq+0)^{3/2}}
\end{eqnarray}
\end{subequations}
This is just the expected (Fourier transform of) the singular 
solution of continuum
elasticity as one approaches the crack tip. Thus, as in all boundary-layer
problems, the large-distance limit of the ``inner'' solution
matches on to the short-distance limit of the ``outer'' solution,
confirming  the justification of the choice of $F_0$ above.
We need to relate the strength of the short-distance singularity of the
continuum solution to the driving displacement $\Delta$.  This
can be done directly by solving the continuum elastic equations.  It
is easier, however, to follow Slepyan and derive this relation 
by energy considerations.
One can calculate \citep{slepyan} the flux of
energy into the boundary-layer, or ``process" zone, obtaining
\begin{equation}
T=\frac{i k \kappa^2 B^2 }{2}v
\end{equation}
Now consider the flux of energy $T_0$
being used to break bonds, which is simply related to $\vee (0)$
\begin{equation}
T_0=\frac{ k\left(B \left.S^+\right|_{ q=\frac{i}{\eta_k v} }\right )^2 }{2}v \ .
\end{equation}
Then,  using the idea that the Griffith's displacement is determined
by exactly the condition that all the energy flux is needed just
to break the bonds, we have
\begin{equation}\label{delta}
\frac{\Delta}{\Delta_{G}}=\sqrt{\frac{T}{T_0}}=\frac{i^{1/2}\kappa}
{ \left.S^+\right|_{ q=\frac{i}{\eta_k v}}}
\end{equation}
Notice that the factor of $\kappa$ guarantees that the result is
invariant with respect to how the decomposition of $S$ is performed.

In order to make contact with other treatments, we consider the standard
case $k=1$ and $\eta_k=\tilde{\eta}=\eta$, in which all bonds are equivalent.  In this
case, the function $S$ has a particularly simple form,
\begin{equation}
S(q)=\frac{h}{r}=\frac{H}{R}
\end{equation}
with $H^2=4 (1 + i q \eta v) \sin^2 \frac{q}{2}-q^2 v^2$ 
and $R^2=4 (1 + i q \eta v) ( \sin^2 \frac{q}{2}+1)-q^2 v^2$. 
We can represent these functions in  term of the products over all their roots 
\begin{eqnarray}\label{H}
H & = & \sqrt{1-v^2}(q+i 0)^{1/2}(q-i 0)^{1/2} \prod_{n,m} 
\left(1-\frac{q}{Q^+_n}\right)^{1/2}
\left(1-\frac{q}{Q^-_m}\right)^{1/2} \\  \label{R}
R &= & 2 \prod_{n,m} 
\left(1-\frac{q}{q^+_n}\right)^{1/2}
\left(1-\frac{q}{q^-_m}\right)^{1/2}
\end{eqnarray}
where $Q^+_n(q^+_n)$ are the (nonzero) roots of $H^2$($R^2$) in 
the lower half plane and 
$Q^-_n(q^-_n)$ are corresponding roots of $H^2$($R^2$) in the 
upper half plane. 
This decomposition allows us to write down explicitly the factors
$S^+$, $S^-$:

\begin{subequations}
\begin{eqnarray}\label{Spl}
S^+(q)=\frac{(1-v^2)^{1/4}}{\sqrt{2}}(q+i 0)^{1/2}\prod_{n,m}\left(\frac{1-
\frac{q}{Q^+_n}}{1-\frac{q}{q^+_m}}\right)^{1/2} \\
S^-(q)=\frac{(1-v^2)^{1/4}}{\sqrt{2}}(q-i 0)^{1/2}\prod_{n,m}\left(\frac{1-
\frac{q}{Q^-_n}}{1-\frac{q}{q^-_m}}\right)^{1/2}
\end{eqnarray}
\end{subequations}
Examining the small $q$ behavior, we find the $\kappa=1$, so that
substitution of $S^+$ into Eqn. (\ref{delta}) gives
\begin{equation}\label{e1}
\frac{\Delta}{\Delta_{G}}=\frac{(2 \eta v)^{1/2}}{(1-v^2)^{1/4}} \prod_{n,m}
\left( \frac{1-\frac{i}{\eta v q_n^+}}{1-\frac{i}{\eta v Q_m^+}}\right)^{1/2}
\end{equation}

If we explicitly take out of the product in Eq. (\ref{e1}) the one imaginary 
$q^+$, which we will denote as $q^+_0$, we get
\begin{equation}
\frac{\Delta}{\Delta_{G}}=\frac{1}{(1-v^2)^{1/4}}\sqrt{
\frac{2(1 + i v \eta q^+_0)}{i q^+_0}} \prod_{n \ne 0 ,m} \left( \frac{
1 + i v \eta q^+_n}{ 1 + i v \eta Q^+_m} \frac{Q^+_m}{q^+_n}\right)^{1/2}
\end{equation} 
This formula was obtained by \citet{dk}, who started from a
finite lattice of transverse size $N_y$  and considered the $N_y \rightarrow
\infty$ limit. Note that this can be done for $k=1$ on a square lattice, 
(the case
being considered here), but not for general $k$ or for modes I or III
on a triangular lattice; the
method here works for these cases as well \citep{plk-new}. 

Next, it is easy to
see that in the limit $\eta \rightarrow 0^+ $, we have 
$\left.S\right|_{ q=\frac{i}{\eta v}}=1$ and that therefore
\begin{equation}
\frac{\Delta}{\Delta_{G}}=\left( i \left.\frac{S^-}{S^+}\right|
_{ q=\frac{i}{\eta v}}\right)^{1/2}=i^{1/2}\prod_{n,m} 
\left( \frac{ q^-_m Q^+_n} {q^+_m Q^-_n} \right)^{1/4} \ .
\end{equation}
Furthermore, in this
limit $q^+_0 = -q^-_0$. All other complex roots cancel out, leaving only
real roots. Hence, 
\begin{equation}
\frac{\Delta}{\Delta_{G}}=
\prod_{over \> real \> Q_n,\, q_m} \left( \frac{ q^-_m Q^+_n} 
{q^+_m Q^-_n} \right)^{1/4} \ ,
\end{equation}
a result originally obtained by Slepyan. Note that in our
treatment, the presence of one of these real roots in either
the numerator or the denominator of the product expression
depends entirely on the half-plane from where it came as we took
the limit  $\eta \rightarrow 0^+ $. To see what this condition implies,
let us denote by  $H^2$(or $R^2$) as $f$; then the small imaginary part
of a specific  real root, $i \delta$, satisfies the equation
\begin{equation}
\left.i\delta\frac{\partial f}{\partial q}
\right|_{\eta=0,\; q=q_{real}}+\eta\left.\frac{\partial f}{\partial \eta}\right|_{\eta=0,\; q=q_{real}}=0  \ ,
\end{equation}
which gives as the determining factor
\begin{equation}
{\delta \over \eta} =\left.\frac{- q^3 v^3}{\frac{\partial}{\partial q}(4 \sin^2
\frac{q}{2}-v^2 q^2)}\right|_{q=q_{real}}\ .
\end{equation}
This condition is exactly what appears in  Slepyan's work.

\section{Numerical calculations}

In this section we derive an integral form for
the basic result Eq. (\ref{e1}). We then present a numerical procedure for
finding the displacement-velocity curve.
The basic notion is to utilize Eq.
(\ref{log}) to find $S^+$. It is  more convenient for  the numerical
work to isolate explicitly the singularity at $q=0$, defining $K(q)$  by
\begin{equation}\label{sandk}
S^2(q)=\frac{A^2 \phi ^2 ( q + i 0)(q-i 0)}{q^2 + \phi ^2} K(q)
\end{equation}
where $\phi$ is an arbitrary positive constant. Because of the factor $A^2 q^2$
in the denominator, $K(0)=1$; the 
factor $(q^2 + \phi^2)/ \phi^2$ does not alter the desired asymptotic behavior
at infinity: $K \rightarrow const$ as $q  \rightarrow \pm \infty$. 
Applying Eq. (\ref{log}), we get for $Im \;q>0$
\begin{equation}\label{spl}
S^+=\exp{ \frac{1}{4 \pi i} \int_{-\infty}^{+\infty}
\frac{\ln S^2(\gamma)}{\gamma-q} d \gamma}=
\left(A \phi \frac{q+i 0}{q+i\phi} \right)^{\frac{1}{2}}\left(K^+ \right)^{\frac{1}{2}}
\end{equation}
and for $Im \;q<0$
\begin{equation}\label{sm}
\frac{1}{S^-}=\exp{ \frac{1}{4 \pi i} \int_{-\infty}^{+\infty}
\frac{\ln S^2(\gamma)}{\gamma-q} d \gamma}=
\left(\frac{1}{A \phi} \frac{q-i \phi}{q-i 0} \right)^{\frac{1}{2}}
\left(K^-\right)^{-\frac{1}{2}}
\end{equation}
with 
\begin{equation}\label{kpm}
\left(K^\pm \right)^{\pm \frac{1}{2}}=\exp{ \frac{1}{4 \pi i} 
\int_{-\infty}^{+\infty}\frac{\ln K(\gamma)}{\gamma-q} d \gamma}
\end{equation}
where the upper sign is for $Im \;q>0$ and the lower sign  for $Im \;q<0$.

There is one point regarding these expressions that must be clarified.
We must make the correct choice of the branch cuts for the square roots in
$h$ and $r$ appearing in $S$. It is convenient to express things in
terms of $S^*(q)\equiv h/r = H/R$, the simple  $k=1$ and $\eta_k=\eta$,
form of $S$, so that from Eqn. (\ref{seqn})
\begin{equation}
\label{seqn1}
S = \frac{S^*(1+i\eta v q)}{S^*(1+i\eta v q) + (1-S^*)k(1 + i\eta_k v q)} \  ,
\end{equation}
and from Eqn. (\ref{xi})
\begin{equation}
\xi=\frac{1-S^*}{1+S^*} \ .
\end{equation}
We must choose $S^*$ such that
$|\xi|<1$ as noted below Eqn. (\ref{xi}). It is easy to see from the
above expression for $\xi$ that it is sufficient to choose
$S^*$ to lie in the first and fourth quadrants to guarantee this.  For
$S^*$, we can then generate $S$ directly via Eqn. (\ref{seqn1}), and from
this $K$.

As $K(0)=1$, there is no pole in the integral (\ref{kpm}) at $\gamma=0$ 
when $q=0$. This means that 
$K^+(0)=(K^-(0))^{-1}$ and the asymptotic expressions for 
$S^{\pm}$ are
\begin{equation}  
S^+|_{q \rightarrow 0}=\sqrt{\frac{A K^+(0)}{i}} ( q + i 0)^
\frac{1}{2};
\; \; \; 
S^-|_{q \rightarrow 0}=\sqrt{\frac{i A}{K^+(0)}} ( q - i 0)^\frac{1}{2}
\ .\end{equation}
Comparing this to Eqn. (\ref{kappa}), we have 
\begin{equation}
\kappa=\sqrt{-iK^+(0)} \ ,
\end{equation}
so that
\begin{equation}
\frac{\Delta}{\Delta_{G}}=\sqrt{\frac{1+\phi \eta_k v}{A  \phi} 
\frac{K^+(0)}{K^+(\frac{i}{\eta_k v})}}
\end{equation}
Explicitly writing out the  integrals,  we get
\begin{equation}\label{numres3}
\frac{\Delta}{\Delta_{G}}=\sqrt{\frac{1+\phi \eta_k v}{A  \phi}}
\exp {\frac{1}{4 \pi i} \int_{-\infty}^{+\infty}\frac{\ln K(\gamma)}
{\gamma(1+ i \eta_k v \gamma) } d \gamma}
\end{equation}
The extra degree of convergence of the integrand at infinity
is very desirable, as it helps control the fact that $\ln K(\xi)$
is a rapidly oscillating function. Also, the fact that the answer must be
independent of $\phi$ provides a check on the numerical routines.
A  change of variables $\gamma \rightarrow - \gamma$
allows the integral to be rewritten as
\begin{equation}
\frac{\Delta}{\Delta_{G}}=\sqrt{\frac{1+\phi \eta_k v}{A  \phi}}
\exp {\frac{1}{4 \pi} \int_{0}^{+\infty} Im \frac{\ln K(\gamma)}
{\gamma(1+ i \eta_k v \gamma) } d \gamma}
\end{equation}
To actually compute this integral, we  divided the 
region of integration into two parts from $0$ to $1$  and from $1$ to
$+\infty$. The second integral was further transformed 
so as to have in the numerator the factor $\ln A^2 \phi^2 K(\xi)$
which goes to zero at infinity; the subtracted integral with the numerator
$\ln 1/(A \phi)^2$ was performed analytically. After all these manipulations,
the integrals were successfully computed by standard mathematical library 
subroutines. 

\begin{figure}\centerline{\includegraphics[width=3.25in]{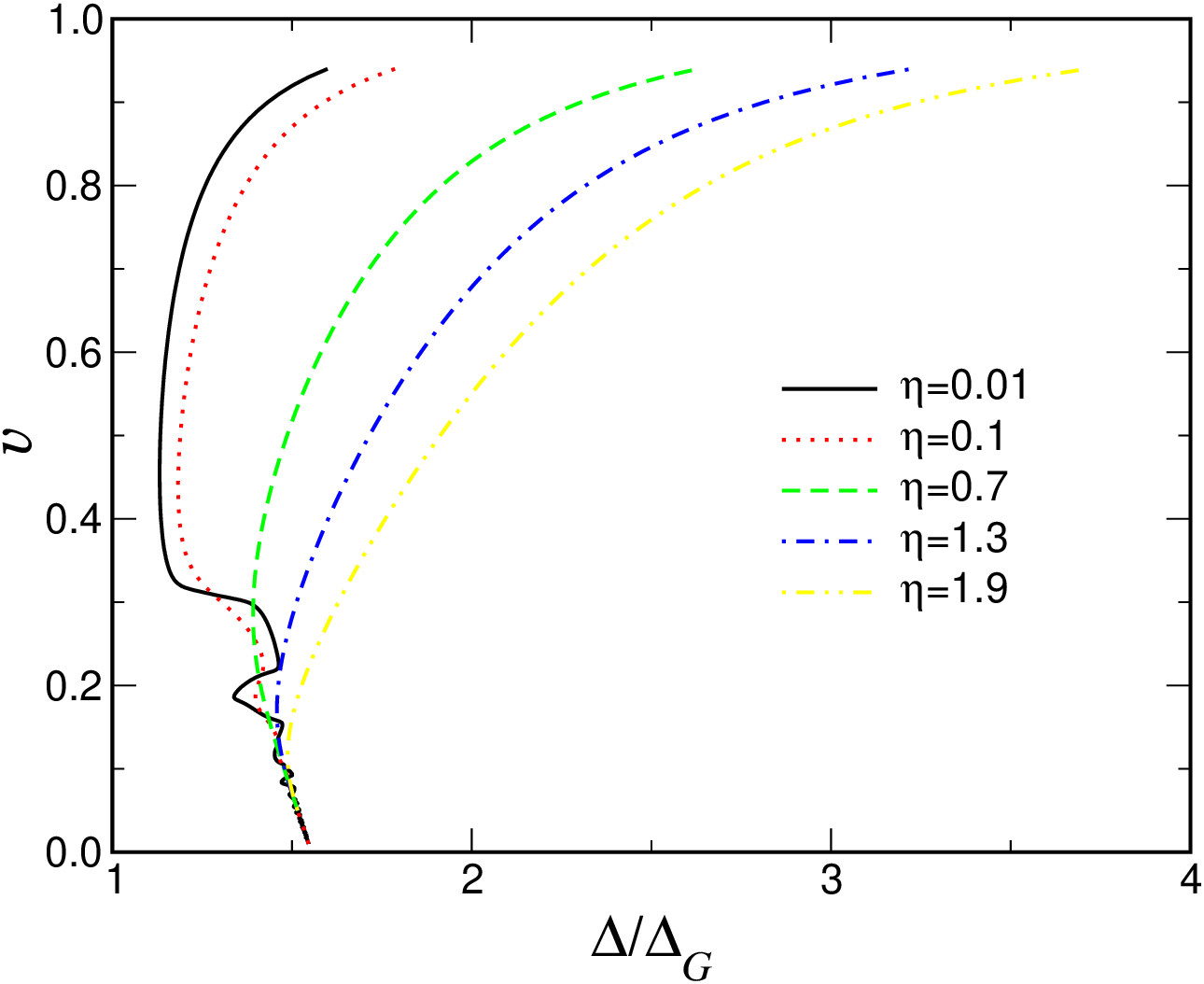}}
\caption{Dependence of speed $v$ on dimensionless driving
$\frac{\Delta}{\Delta_G}$ for $k=1$ and $\eta_k=\eta$. The curves are 
for $\eta=0.01$, $0.1$, $0.7$, $1.3$, $1.9$}
\label{figure1}
\end{figure}

\begin{figure}\centerline{\includegraphics[width=3.25in]{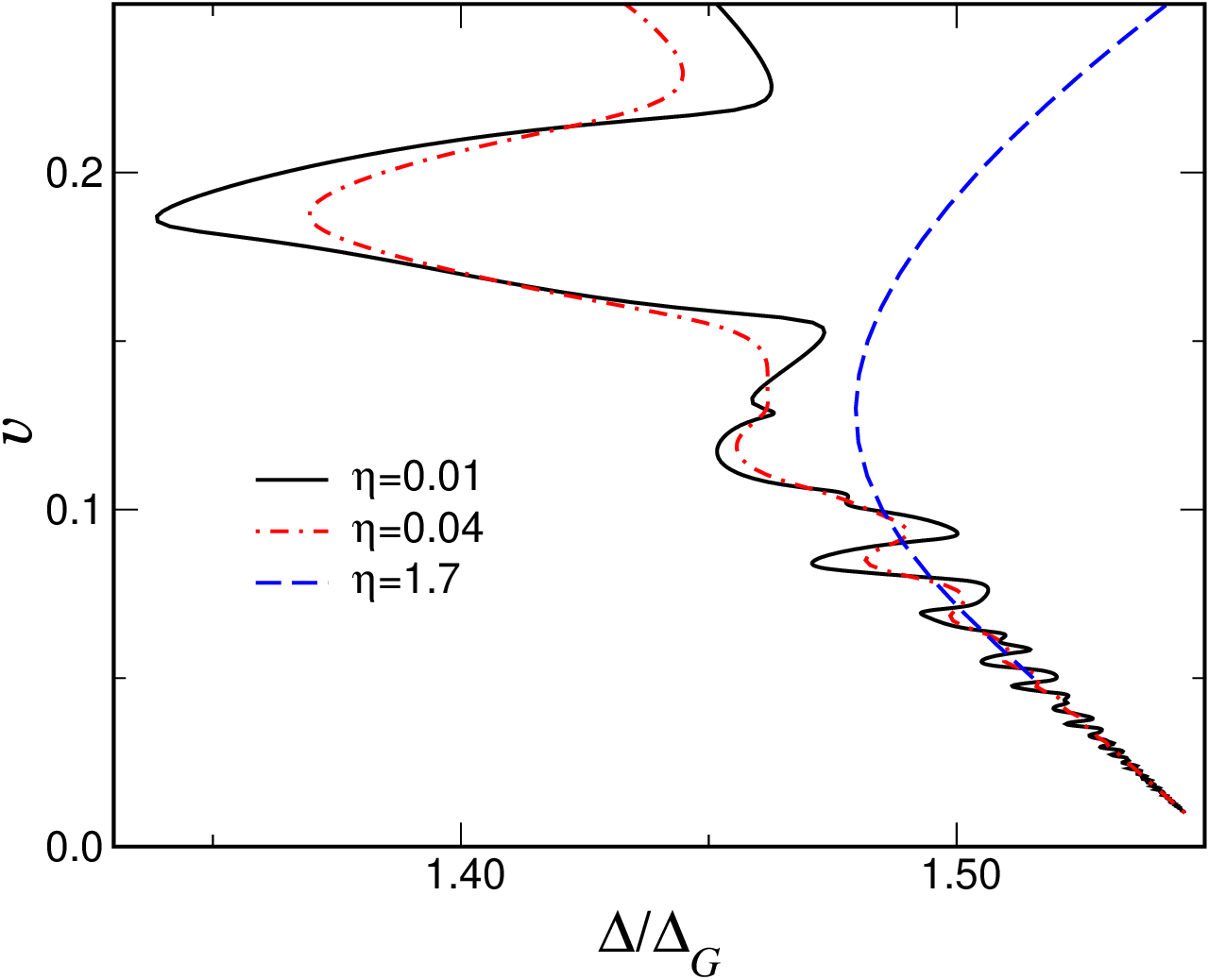}}
\caption{Change from  
rapid oscillations for small $\eta$ to smooth behavior at large  $\eta$.
Here $k=1$ and $\eta_k=\eta$.}
\label{figure2}
\end{figure}

\begin{figure}\centerline{\includegraphics[width=3.25in]{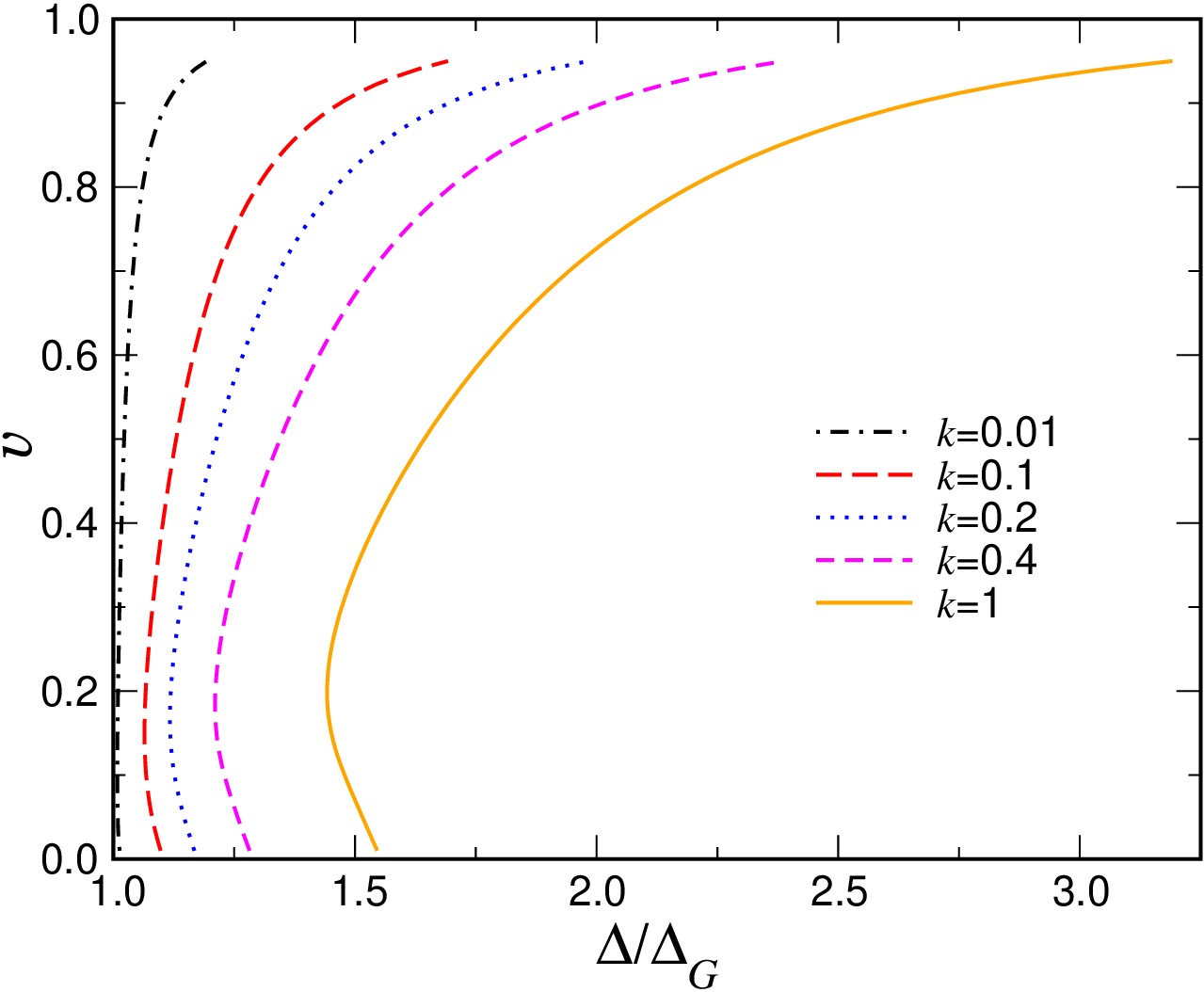}}
\caption{Dependence of speed $v$ on dimensionless driving
$\Delta/\Delta_G$ for $k=0.01$, $0.1$, $0.2$, $0.4$, $1.0$. Here
$\eta=\eta_k=1.1$}
\label{vsmallk}
\end{figure}

Results from our numerical calculations for $k=1$ and $\eta_k=\tilde{\eta}=\eta$, 
are presented in
Figs. \ref{figure1},  \ref{figure2}, which reproduce, of course,
those of \citep{dk}. At small $\eta$, there are large
oscillations in the $v$ vs $\Delta$ curve, as found initially by
Marder and Gross for lattices with small transverse sizes. The
curves all hit the $v=0$ line at the point which marks the end of the
band of lattice-trapped static cracks \citep{kl1}. For larger damping,
the oscillations disappear and the moving crack branch bifurcates
smoothly (in the backward direction) from this point, eventually turning
around and becoming stable. In Fig. \ref{vsmallk}, we contrast
the standard result to that for smaller $k$, keeping $\eta_k=\eta$.
We see that the $v=0$ limit has moved closer to unity, in accord with
the finding in \citep{kl1} that the window of arrested cracks 
narrows
as $k$ decreases (note for purposes of comparison that our $k$ is a factor
of 2 smaller than that defined in \citep{kl1}.  The velocity
at the minimum $\Delta/\Delta_G$ however is hardly affected, so the
minimum stable velocity is essentially unchanged.  Past this point,
 the velocity is higher at smaller $k$, even when the change in $\Delta_G$
is factored in.  This is further evidence of how the microscopic details
influence the velocity versus driving displacement curve.

\section{Consistency of the solution}

As mentioned in the introduction, cracks become unstable above a critical
velocity. In the framework of the model being considered here, one
might assume that this velocity is associated with the solution as
found by the Wiener-Hopf method becoming inconsistent. As we shall
see, there  are two kinds of inconsistencies.  The first sets
in for small $v$, for small enough $\eta$.  Here the vertical bond
between $y=0$ and $y=1$ first achieves critical extension prematurely,
at some positive $\tau$, contradicting the assumption that the bond breaks
at $\tau=0$. In the second kind of inconsistency, a  horizontal
bond achieves critical extension.  If the model is defined so that
only the vertical bonds between $y=0$ and $y=1$ are breakable, then
this does not present a problem.  If one the other hand, all the
springs are chosen identical, so that the horizontal springs also
break at the same critical extension as the vertical ones, then the
solution is indeed inconsistent at this point.  This is what occurs
at large enough velocity, for all $\eta$.

To proceed, we must calculate the bond lengths, back transforming our Fourier
space solutions so as to obtain the physical displacements. 
We find the time dependence of the function $\vee(\tau)$.
From Eqs. (\ref{vp}), (\ref{vm}) and  (\ref{v0bk}),  we have 
\begin{equation}
\vee^+=\frac{\vee (0)}{(-iq+0)(1 + i q \eta_k v)}\frac{S^+}{\left.S^+ (q)
\right| _{q=\frac{i}{\eta_k v}}}-\frac{ \eta_k v \vee (0)}{1 + i q \eta_k v}\ ;
\end{equation}
\begin{equation}
\vee^-= \frac{ \vee (0)}{\left.S^+ 
\right| _{q=\frac{i}{\eta_k v}}S^-(q) }\frac{1}{(iq+0)(1 + i q \eta_k v)}
+\frac{ \eta_k v \vee (0)}{1 + i q \eta_k v} \ .
\end{equation}
Performing the inverse Fourier transform, we get
\begin{equation}\label{vpos}
\vee(\tau)=\vee(0)\int_{-\infty}^{+\infty}\frac{ \sqrt{ A \phi} 
e^{- i q \tau}  }
{\sqrt{(i q - 0)( i q - \phi)}  ( 1 + i q \eta_k v )}\frac{(K^+)^
{\frac{1}{2}} }{\left.S^+ 
\right| _{q=\frac{i}{\eta_k v}}  }\; \frac{d \, q}{2 \pi}
\end{equation} for $\tau>0$ and 
\begin{equation}\label{vneg}
\vee(\tau)=
\vee(0) \int_{-\infty}^{+\infty}\frac{ e^{- i q \tau}}
{\sqrt{A \phi} (i q + 0 )( 1 + i q \eta_k v )}\left(\frac{i q +\phi}{
iq+0}\right)^{\frac{1}{2}} \frac{(K^-)^
{-\frac{1}{2}} }{\left.S^+ 
\right| _{q=\frac{i}{\eta_k v}}  }\; \frac{d \, q}{2 \pi} +
\vee(0) e^{\frac{\tau}{\eta_k v}}
\end{equation}
for $\tau<0$.
Here we have used the expressions for $S^{\pm}$ derived in the
last section.

Clearly, these integrals must be done numerically. To proceed, we
transform the pieces containing the factors $K^{\pm}$ by dividing both the
numerator and denominator by $\left(K^+(0)\right)^{1/2}$. In the denominator,
we get $\Delta_{G}/\Delta$; we have already discussed how this 
can be computed.  
We now have integrands containing  factors of the form
\begin{equation}
\frac{\left(K^\pm(q)\right)^{\pm \frac{1}{2}}}
{\left(K^+(0)\right)^{\frac{1}{2}}}=\exp \frac{1}{4 \pi i}
\int_{-\infty}^{+\infty}\frac{q \ln K(\xi)}{\xi (\xi-q)} d \xi \ .
\end{equation}
The $\pm$ refers on the right hand side to the sign of the imaginary part 
of $q$. As we will see later on,
we need to calculate this function only for positive $q$.
Then, the integral in the exponent can be written as
\begin{equation}
\frac{1}{2 \pi i}
\int_{0}^{+\infty}\frac{ Re \ln K(q \xi)}{\xi^2-(1\pm i 0)^2} d \xi +
\frac{1}{2 \pi}
\int_{0}^{+\infty}\frac{ Im \ln K(q \xi)}{\xi (\xi^2-(1\pm i 0)^2)} d \xi \ .
\end{equation}
We break up the region of integrations into three parts,$(0,1/2)$, $(1/2, 3/2)$
and $(3/2, +\infty)$. In the first interval, we numerically calculate the integral 
directly as written. For the third interval,  we proceed as discussed 
in the last section for a similar 
semi-infinite integral; this leaves us with having to integrate 
numerically a function which behaves
as $1/\xi^4$ near infinity. Finally, for the integral near $1$
we add an integral with $K (q \xi)$ replaced by $K (q)$;
this added integral is done analytically and the resultant
subtracted integral with the integrand now containing $\ln K(q \xi)/K(q)$
is done numerically. 

Applying the described procedure, we can evaluate the functions 
$\left(K^\pm(q)\right)^{\pm \frac{1}{2}}/\left(K^+(0)\right)^{\frac{1}{2}}$ 
for any real positive $q$. We must now perform the final integration
over the variable $q$.  First, we note that the requirement of having
a real displacement necessitates $\vee^\pm(-q)=\vee^\pm
(q)^*$. This allows us to change the region of integration to $(0, +\infty)$
in (\ref{vpos}); we will return shortly to \ref{vneg}. The trickiest part of this calculation
concerns the behavior near $q=0$. For $\vee^+$, there is a $1/q^{1/2}$
behavior. Here the leading order term can be integrated analytically
over the  interval $(0,1)$ and 
then the overall function with this term subtracted can be used for a 
numerical 
integration. This works in a  straightforward manner.

We need to be more careful with the integral (\ref{vneg}), because 
here the integrand 
behaves near zero as $1/q^{3/2}$. To proceed, we first 
divide the interval of integration into three parts: 
$(-\infty, -1)$, $(-1, 1)$,
and $(1, +\infty)$. The integrals over $(-\infty, -1)$  and 
$(1, +\infty)$ can again be combined to yield an integral twice the 
real part of the integrand over the latter range.
For the range spanning zero, we first subtract from the
integrand the leading term 
$$\frac{1}{\sqrt{A}\left.S^+ \right| _{q=\frac{i}{\eta_k v}}
(iq+0)^{3/2}} \ .$$
After this subtraction, the  integrand becomes of  order
$1/q^{1/2}$. Now, the one-sided integral converges, and thus we can
again transform the range to $(0, 1)$. This subtracted integral is
evaluated numerically from a very small value of $q$ (typically
$q=10^{-3}$) to $1$. In the remaining part of the range,
i.e. from $0$ to $10^{-3}$, we evaluate the integral analytically 
after approximating it by its leading small $q$ behavior,  
$\alpha/q^{1/2}$; the value of the constant $\alpha$
is  determined by fitting the the behavior of the function near the
point $10^{-3}$.
Finally we need to perform analytically the integral of
the subtracted piece   $1/(\sqrt{A}\left.S^+ \right| _{q=\frac{i}{\eta_k v}})
/(iq+0)^{3/2}$. To accomplish this,  we deform the contour of integration 
from $(-1,1)$ on the real axis
to the curve from $-1$ to $1$ going over the lower half of the unit circle.
Note that the branch-cut for this function must be taken as before over negative 
real axis; this choice guarantees the fact that  $f(-q)=f^*(q)$. Note that
once all our transformations are complete, we only need values for
the integrand for positive (real) values of $q$; this was used
earlier in the technique for calculating $K^\pm$.
A good check of this 
complex numerical technique of computing the
Fourier transformation is applying it to the Fourier transform 
$\vee^+(q)$ for $\tau<0$ and $\vee^-(q)$ for $\tau>0$,
as in this case result must be equal to zero.

\begin{figure}\centerline{\includegraphics[width=3.25in]{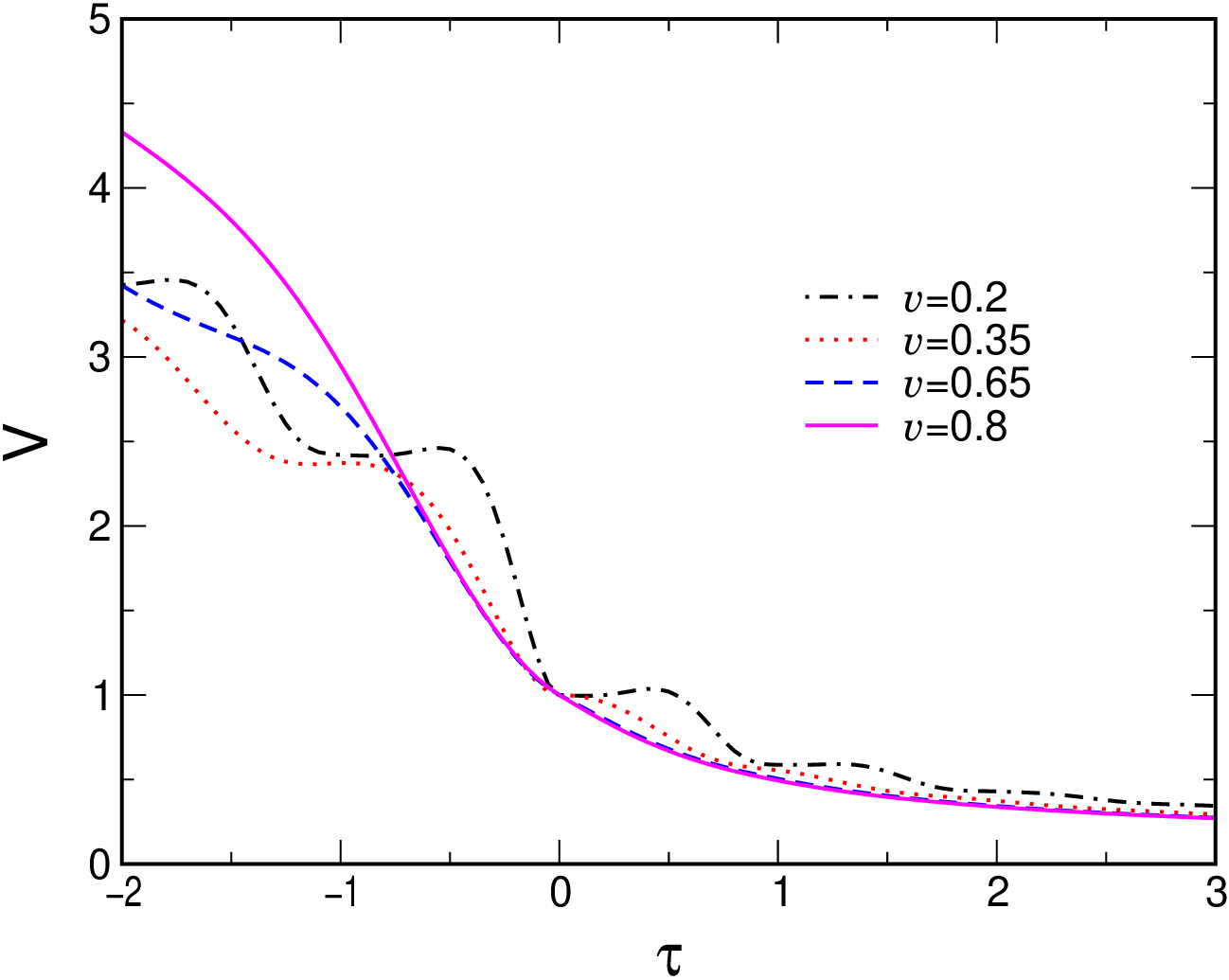}}
\caption{Elongation of vertical bonds along the path of the crack
for $\eta=0.2$. Here $k=1$ and $\eta_k=\eta$}
\label{figure3}
\end{figure}

We first examine what happens in the standard case, $k=1$ and $\eta_k=\tilde{\eta}=\eta$.
Figure \ref{figure3} shows the behavior of $\vee(\tau)$ for $\eta=.2$ and
various values of the velocity. As discussed above, we see the two
different kinds of inconsistencies exhibited by these solutions. 
For example, the solution for $v=0.2$ is inconsistent as 
$\vee(\tau)/\vee(0)>1$ for $\tau>0$.
This type of inconsistency, seen at small velocities,  corresponds to
the region of backward behavior on $v-\Delta$ graph. As
already claimed by Marder and Gross, this region is therefore unphysical.

\begin{figure}\centerline{\includegraphics[width=3.25in]{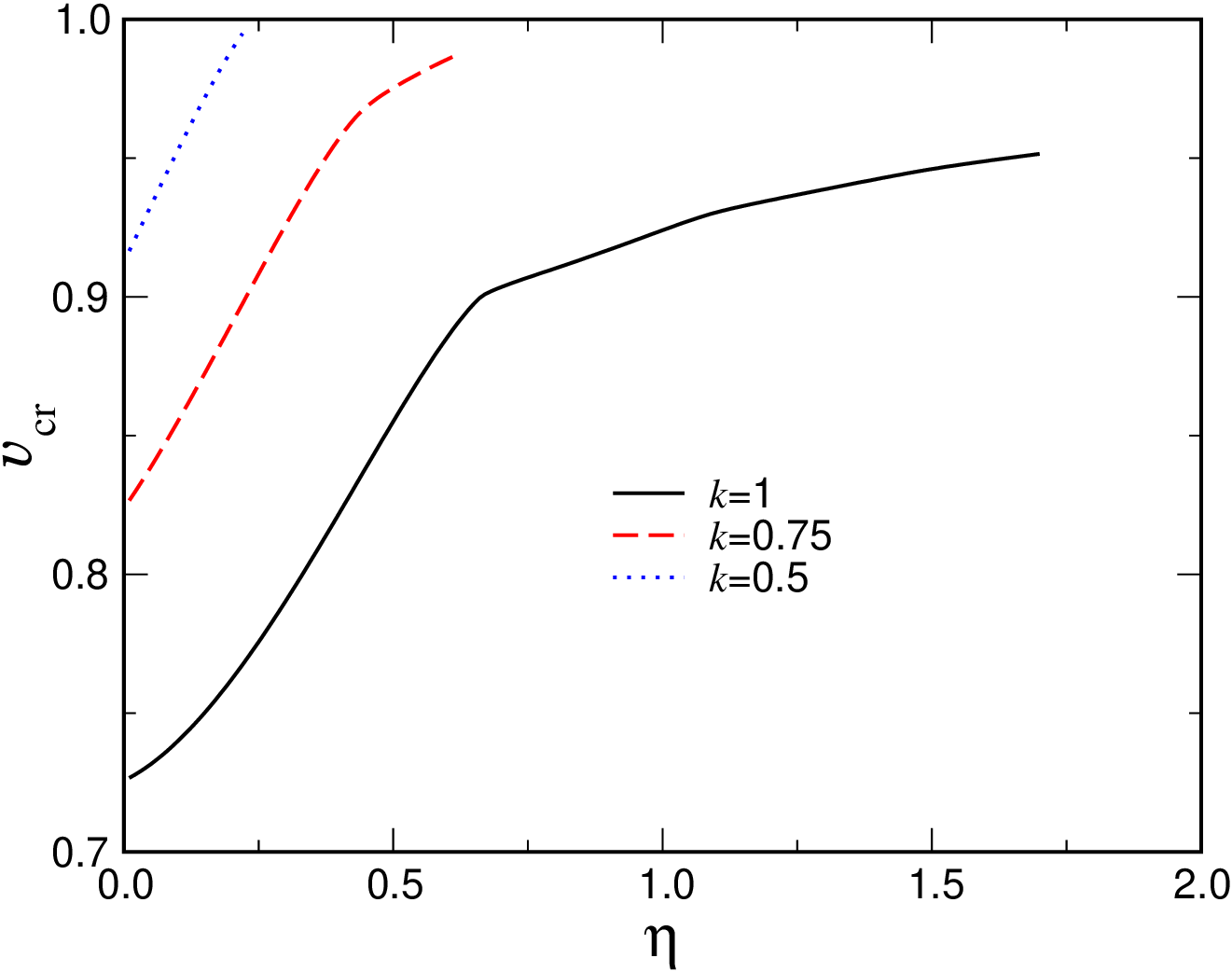}}
\caption{Dependence of the critical speed $v_{cr}$ on $\eta$ for
$k=1$,  $0.75$, $0.5$.  Here $\eta_k=\eta$}
\label{figure4}
\end{figure}

\begin{figure}\centerline{\includegraphics[width=3.25in]{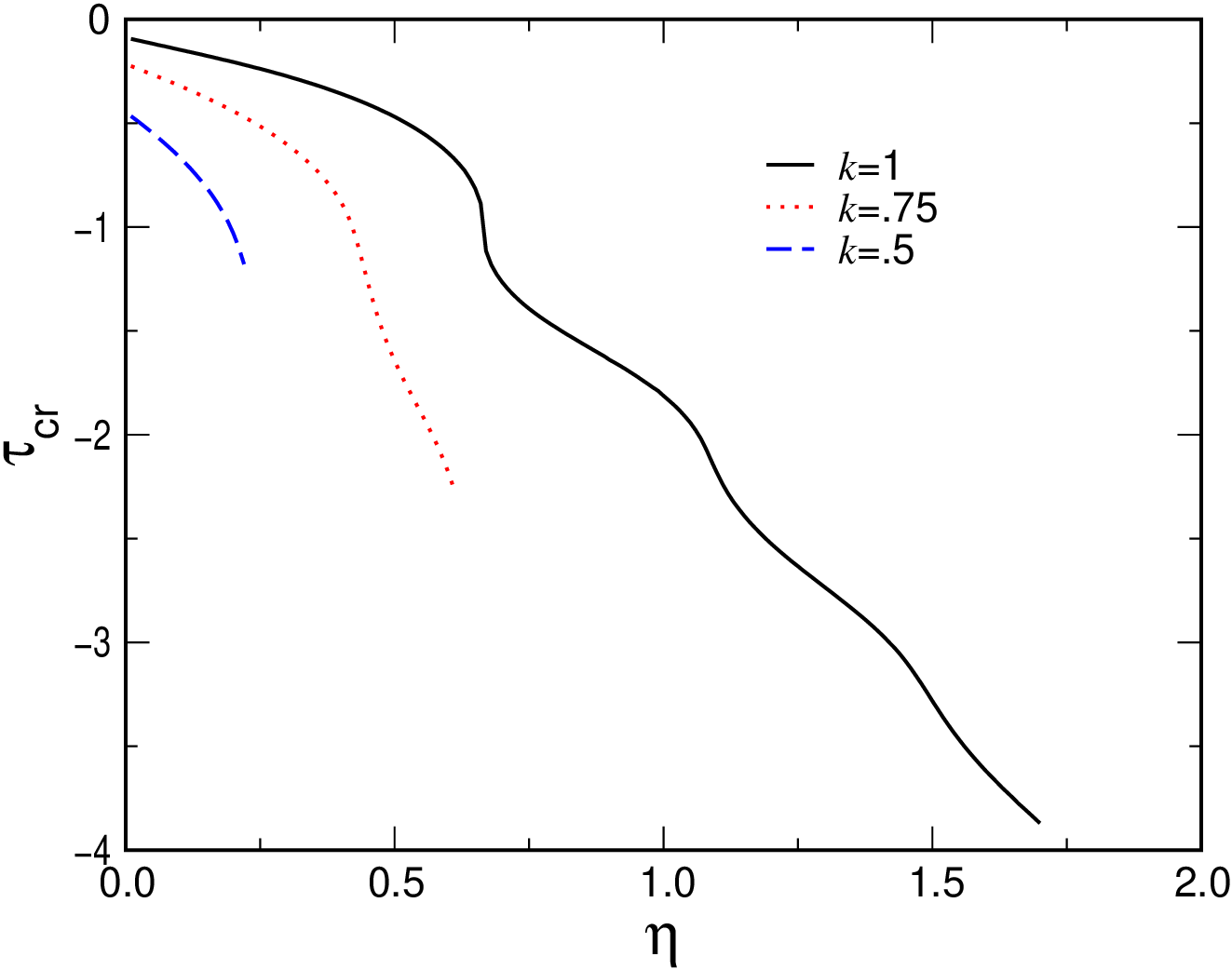}}
\caption{Dependence of the critical time $\tau_{cr}$ on $\eta$ for $k=1$,
$0.75$, $0.5$. Here $\eta_k=\eta$.}
\label{figure5}
\end{figure}

More interesting is the inconsistency which sets in at large
velocity, and has been argued to be related to the experimentally
observed microbranching. What happens here is that the elongation 
of the horizontal bond, given by $\vee_{hor}(\tau)=(\vee(\tau-1)-\vee(\tau))/2$
increases as a function of increasing driving. 
At some critical speed it reaches the value $\vee(0)$ and the bond breaks.
Figure \ref{figure4} shows the dependence of the critical speed on $\eta$. 
We note in passing that these
graphs agree with what can be gotten by extrapolating a similar
graph for the finite $N_y$ lattice, a graph that can be obtained
using the methods of \citep{kl1}. The actual location in $\tau$ at which the
horizontal break occurs at the critical velocity is shown in Fig. \ref{figure5}.

Several details of our findings are worthy of note. First, there is always a critical speed
for any dissipation, but this speed gets very close to the
maximal crack speed for large $\eta$. The value of the critical speed
for the dissipation-less limit is $v_{cr} \simeq .725$; this is above
what has been seen in some experiments but recall that we are doing mode III
on a square lattice, a far cry from mode I in an amorphous system. The
is a curious break in the curve at $\eta \simeq .665$, which corresponds in
Fig. \ref{figure5} to the point where the break location varies most
strongly with $\eta$. 

It should be noted that this inconsistency does not appear related in
any obvious way to the \citet{yoffe} criterion.  First, the Yoffe
criterion, concerning the direction of maximal stress, is calculated
for the continuum elastic field, and is thus completely independent of
the viscosity $\eta$ \citep{kl1}.  The onset of the inconsistency is strongly
$\eta$ dependent however.  Also, the Yoffe criterion is stated for
Mode I cracks, and its applicability to Mode III is subject to question.

We turn now to the case of $k<1$, with weakened bonds along the crack path,
again keeping $\eta_k=\eta$.
Figures \ref{figure4}, \ref{figure5} show our results.
As is reasonable, the inconsistencies persist, but are pushed  to
higher speeds as $k$ decreases.  It is interesting to speculate that
since for finite width systems, velocities greater than the wave
speed are possible, sufficiently small $k$ might enable stable steady-state
fracture at supersonic velocities. Also, intersonic wave speeds are possible
in the case of mode II fracture, and therefore weakened bonds on the path of the 
crack would in this case as well allow steady-state propagation at higher speeds \citep{science}.

\section{Discussion}

We have shown how to extend the Slepyan approach to cracks in
ideally brittle infinite lattice systems to the case of having a 
Kelvin viscosity for each lattice spring. 
In addition, we have examined the onset of the additional, off-axis
cracking at high velocity.
We showed how dissipation delays but
does not eliminate the microbranching instability, an
effect impossible to see  within the context of any continuum
elastic treatment.

In the following work, we will present the results of similar calculations for
mode I cracks on a triangular lattice. That system is much closer to
those that have been studied experimentally and one can therefore hope for
more direct lessons to emerge. Also, we should mention complementary
studies on nonlinear lattice models which relax the assumption of
ideally brittle springs and thereby allow one to view the onset of
inconsistent behavior as the limiting case of more standard bifurcations
\citep{arrest,klnew}. 

\begin{acknowledgments}
DAK acknowledges the support of the Israel Science Foundation.  The
work of HL and LP is supported in part by the NSF, grant no. DMR94-15460. DAK
and LP thank Prof. A. Chorin and the Lawrence Berkeley National Laboratory
for their hospitality during the inital phase of this work. 
\end{acknowledgments}

\section{Appendix}
We  discuss here some of our conventions and notations
regarding Fourier transformations.
We use Fourier transformation in the form
\begin{eqnarray} \label{Fourier}
u^F(q) =\int_{-\infty}^{+\infty}u(\tau)e^{i q \tau}d \tau \nonumber \\
u(\tau)=\frac{1}{2\pi}\int_{-\infty}^{+\infty}u^F(q)e^{-i q \tau}dq
\end{eqnarray}
We define also $u^+$ and $u^-$ as
\begin{subequations}
\begin{eqnarray}
u^+(q) &=& \int_{0}^{+\infty}u(\tau)e^{i q \tau}d \tau \\
u^-(q) &=& \int_{-\infty}^{0}u(\tau)e^{i q \tau}d \tau 
\end{eqnarray}
\end{subequations}
This gives 
\begin{subequations}
\begin{eqnarray}
u(\tau)\theta(\tau) &=& \frac{1}{2\pi}\int_{-\infty}^{+\infty}u^+(q)e^{-i q \tau}dq \ ;\\
u(\tau)\theta(-\tau) &=& \frac{1}{2\pi}\int_{-\infty}^{+\infty}u^-(q)e^{-i q \tau}dq \ ,
\end{eqnarray}
\end{subequations}
which means that $u^+$ has poles only in the lower half plane
and $u^-$ only in the upper half plane. 

Now let us define $p=-i q$. Then
\begin{equation}
\int_{0}^{+\infty}u(\tau)e^{i q \tau}d \tau \stackrel{_{p\rightarrow+\infty}}{\sim} u(0) \int_{0}^{+\infty}e^{-p \tau}d \tau
=\frac{1}{p} u(0)\ .
\end{equation}
Thus 
\begin{equation}\label{limp}
\lim_{- i q \rightarrow +\infty}{-i q u^+}=u(0) \ .
\end{equation}
Similarly we can get for $p=i q$ 
\begin{equation}
\int_{-\infty}^{0}u(\tau)e^{i q \tau}d \tau \stackrel{_{p\rightarrow+\infty}}{\sim} u(0) \int_{-\infty}^{0}e^{p \tau}d \tau
=\frac{1}{p} u(0) \ .
\end{equation}
So 
\begin{equation}\label{limm}  
\lim_{i q\rightarrow+\infty}{i q u^-}=u(0) \ .
\end{equation}

The Fourier transform of $\theta(-\tau)\frac{d}{d\tau}u(\tau)$ is given by
\begin{equation}\int_{-\infty}^0\frac{du(\tau)}{d\tau}e^{ i q \tau}d\tau
=u(0)-i q u^-\label{derivative} \end{equation}
Finally, we  consider separating an arbitrary function $S(q)$ into
the product of two pieces  
$S^+$ and $S^-$,
with poles and zeroes in lower and upper half-planes respectively.
These can be done using the identity, 
\begin{eqnarray}\label{log}
\frac{1}{2 \pi i}\int_{-\infty}^{+ \infty}\frac{\ln S(\xi)}{\xi-q} d\xi=
\frac{1}{2 \pi i}\int_{-\infty}^{+ \infty}\left(\frac{\ln S^+(\xi)}{\xi-q}+
\frac{\ln S^-(\xi)}{\xi-q}\right)d\xi \nonumber \\
\ =\ \ \ \ \ \left\{
\begin{array}{ll}
for \;\;  Im \;q>0, &\;\;\ln S^+(q)\\
for \;\;  Im \;q<0, &\;-\ln S^-(q)
\end{array}
\right. \ .
\end{eqnarray}


\begin{thebibliography}{99}

\bibitem[Abraham, et al.(1994)]{md0}Abraham, F.F., Brodbeck, D., Rafey, R.A.,
Rudge, W.E., 1994. Instability dynamics of fracture - a computer-simulation investigation. \prl {\bf 73} (2), 272-275.
\bibitem[Barenblatt(1959)]{barenblatt} Barenblatt, G.I., 1959. 
The formation of equilibrium
cracks during brittle fracture:  General ideas and hypothesis, axially 
symmetric cracks. Appl. Math. and Mech. {\bf 23}, 622-636
(1959).
\bibitem[Fineberg, et al.(1991)]{exp1} Fineberg, J., Gross, S.P., Marder, 
M., Swinney, H.L., 1991. Instability in dynamic fracture. \prl {\bf 67}, (4) 
457-460.
\bibitem[Fineberg, et al.(1992)]{exp2} Fineberg, J., Gross, S.P., Marder, 
M., Swinney, H.L., 1992. Instability in the propagation of fast cracks. 
\prb {\bf 45}, (10) 5146-5154.
\bibitem[Fineberg and Marder(1999)]{review} Fineberg, J., Marder, M., 1999. 
Instability in dynamic fracture. Phys. Repts. {\bf 313}, (1-2) 1-108.
\bibitem[Gumbsch, et al.(1997)]{md1a} Gumbsch, P., Zhou, S.J., Holian, B.L., 
1997. Molecular dynamics investigation of dynamic crack stability. 
\prb {\bf 55} (6), 3445-3455. 
\bibitem[Holland and Marder(1997)]{md2}Holland, D., Marder, M., 1997. 
Ideal brittle fracture of silicon studied with molecular dynamics. 
\prl {\bf 80}, (4) 746-749. 
\bibitem[Holland and Marder(1998)]{md2a}Holland, D., Marder, M., 1998.
Erratum: Ideal brittle fracture of silicon studied with molecular dynamics. 
\prl {\bf 81}, (18) 4029.
\bibitem[Kessler(2000)]{dk}Kessler, D.A., 2000.  Steady-state cracks in viscoelastic lattice models II. \pre, to appear.
\bibitem[Kessler and Levine(1998)]{kl1}Kessler, D.A., Levine, H., 1998.
 Steady-state cracks in viscoelastic lattice models. \pre {\bf 59} (5),
 5154-5164.
\bibitem[Kessler and Levine(1999)]{arrest}Kessler, D.A., Levine, H., 1999.
Arrested cracks in
nonlinear lattice models of brittle fracture. \pre {\bf 60} (6)
7569-7571.
\bibitem[Kessler and Levine(2000)]{klnew}Kessler, D.A., Levine, H., 2000.
Stability of Cracks in Nonlinear Viscoelastic Lattice Models. In preparation.
\bibitem[Kulamekhtova(1984)]{slepyan2}Kulamekhtova, Sh.A., Saraikin, V.A., 
Slepyan, L.I., 1984. Plane problem of a crack in a lattice. Izv. AN SSSR. 
Mekhanika Tverdogo Tela, {\bf 19}, (3) 112-118 [Mech. Solids {\bf 19}, (3) 
102-108].
\bibitem[Langer(1992)]{langer} Langer, J.S., 1992. Models of crack 
propagation. \pre {\bf46} (6), 3123-3131.
\bibitem[Langer and Lobkovsky(1998)]{langer-new} Langer, J.S., Lobkovsky, 
A.E., 1998. Critical examination of cohesive-zone models in the theory of 
dynamic fracture. J. Mech. Phys. Solids {\bf 46} (9), 1521-1556.
\bibitem[Marder and Gross(1995)]{marder_gross}Marder, M., Gross, S.P., 1995.
Origin of crack tip instabilities. J. Mech. Phys. Solids {\bf 43} (1), 1-48.
\bibitem[Marder and Liu(1993)]{marder-liu} Marder, M., Liu, X., 1993. 
Instability in lattice fracture. \prl {\bf 71} (15),
2417-2420.
\bibitem[Omeltchenko, et al.(1997)]{md3}Omeltchenko, A., Yu, J.,  Kalia, 
R.K., Vashishta, P., 1997. Crack front propagation and fracture in a graphite 
sheet: a molecular-dynamics study on parallel computers. \prl, 
{\bf 78} (11), 2148-2151.
\bibitem[Pechenik, et al.(2000)]{plk-new}Pechenik, L., Levine, H., Kessler, D.A., 2000. Steady-state mode I cracks in a viscoelastic  triangular lattice.
Submitted for publication
\bibitem[Pla, et al.(1998)]{sander1} Pla, O., Guinea, F., Louis, E.,Ghasias,
S.V., Sander, L.M., 1998. Viscous effects in brittle fracture. 
\prb {\bf 57} (22), R13981-R13984.
\bibitem[Rosakis, et al.(1999)]{science}Rosakis, A.J., Samudrala, O., Coker, D., 1999.
Cracks faster than the shear wave speed. Science {\bf 284} (5418),
1337-40.
\bibitem[Sander and Ghasias(1999)]{sander2}Sander, L.M., Ghasias, S.V., 1999.
Thermal noise and the branching threshold in brittle fracture. 
\prl {\bf 83}, (10) 1994-1997.
\bibitem[Sharon, et al.(1995)]{exp3} Sharon, E., Gross, S.P. and Fineberg, 
J., 1996. Local crack branching as a mechanism for instability in dynamic fracture.  \prl {\bf 74} (25), 5096-5099.
\bibitem[Sharon, et al.(1996)]{exp4} Sharon, E., Gross, S.P. and Fineberg, 
J., 1996. Energy dissipation in dynamic fracture. \prl {\bf 76} (12), 
2117-2120.
\bibitem[Slepyan(1981)]{slepyan}Slepyan, L.I., 1981. Dynamics of a crack in 
a lattice. Doklady Akademii Nauk SSSR {\bf 258} (1-3) 561-564. 
[Sov. Phys. Dokl. {\bf 26} (5), 538-540].
\bibitem[Slepyan(1982)]{slepyan3}Slepyan, L.I., 1982.The relation between the 
solutions of mixed dynamical problems for a
continuous elastic medium and a lattice. Doklady Akademii Nauk SSSR, 
{\bf 266} (1-3) 581-584. [Sov. Phys. Dokl. {\bf 27} (9), 771-772.
\bibitem[Yoffe(1951)]{yoffe} Yoffe, E.Y., 1951. Philos. Mag. {\bf 42}, 
739-750.
\bibitem[Zhou, et al.(1997)]{md1}Zhou, S.J., Beazley, D.M., Lomdahl, P.S.,
Holian, B.L., 1997. Large-scale molecular dynamics simulations of
three-dimensional ductile failure. \prl
{\bf 78} (3), (479-482).
\end{thebibliography}
\end{document}